\documentclass[showpacs, showkeys,twocolumn,preprintnumbers,floatfix,amsmath,amssymb]{revtex4}
\usepackage{booktabs}
\usepackage{natbib}
\usepackage{amsmath,amsthm}
\usepackage{float}
\usepackage{graphicx}
\usepackage{epstopdf}
\usepackage{caption}
\usepackage{subfigure}
\usepackage{tabularx}
\usepackage{makecell}
\usepackage{threeparttable}
\usepackage{dcolumn}
\usepackage{bm}
\usepackage{color,epsfig,multirow}
\usepackage{array}
\usepackage{hyperref}
\usepackage{algorithm}
\usepackage{algpseudocode}
\usepackage{threeparttable}

\makeatletter
\def\@fnsymbol#1{\ensuremath{\ifcase#1\or *\or \ddagger\or \mathsection\or \mathparagraph\or \|\or **\or \dagger\dagger \or \ddagger\ddagger \else\@ctrerr\fi}}
\makeatother

\theoremstyle{definition}

\theoremstyle{remark}

\allowdisplaybreaks[4]

\begin{document}
\preprint{Preprint}

\title{Spectral convergence of sum-of-Gaussians tensor neural networks\\ for many-electron Schr\"odinger equation}
\author{Teng Wu$^{1,\dag}$}
\author{Qi Zhou$^{2,\dag}$}
\author{Huangjie Zheng$^{2}$}
\author{Hehu Xie$^{3,4}$}\thanks{hhxie@lsec.cc.ac.cn}
\author{Zhenli Xu$^{2,5}$}\thanks{xuzl@sjtu.edu.cn}

\affiliation{$^1$\mbox{Zhiyuan College, Shanghai Jiao Tong University, Shanghai 200240, China.}} 
\affiliation{$^2$\mbox{School of Mathematical Sciences, Shanghai Jiao Tong University, Shanghai 200240, China.}}
\affiliation{$^3$SKLMS, NCMIS, Institute of Computational Mathematics, Academy of Mathematics and Systems
Science, Chinese Academy of Sciences, Beijing 100190, China.}
\affiliation{$^4$\mbox{School of Mathematical Sciences, University of Chinese Academy of Sciences, Beijing 100049, China.}} 
\affiliation{$^5$\mbox{SOG AI-Technology Co. Ltd., Shanghai, China.}}
\affiliation{$^\dag$These authors contributed equally to this work.}

\date{\today}
\begin{abstract}
We present an improved version of the sum-of-Gaussians tensor neural network (SOG-TNN) architecture for solving many-electron Schr\"{o}dinger equation for one-dimensional soft-Coulomb systems.
Model reduction techniques are introduced to reduce the number of tensor-factorized bases under the SOG approximation of the kernel. The Slater determinant ansatz
is employed so that the anti-symmetric property of the wave function can be strictly preserved. Numerical results show that the SOG-TNN achieves 
high accuracy with remarkably small basis sizes. Robust spectral convergence with respect to the basis size is also observed, consistently characterized by 
a mixed algebraic-exponential model for the error decay. These findings validate that the SOG-TNN architecture 
provides an ultra-efficient and low-rank representation of complex multi-electron wave functions, shedding light on  high-fidelity quantum calculations 
in larger-scale many-electron systems.
\end{abstract}




\maketitle

 \section{Introduction}
 \label{sec::intro}

Finding an accurate solution to the many-electron Schr\"{o}dinger equation represents the fundamental challenge of electronic structure theory \cite{Helgaker2014}. The complexity of this problem primarily stems from the curse of dimensionality in high-dimensional integrals \cite{Kohn1999}, further compounded by the necessity of adhering to the Pauli exclusion principle \cite{Pauli1925} and accurately capturing intricate electron correlation effects \cite{Lowdin1955}. Traditional high-precision deterministic solvers, such as full configuration interaction \cite{Knowles1984, Olsen1988} and coupled-cluster theories \cite{Purvis1982, Raghavachari1989, Bartlett2007}, have set the gold standard for precision via linear expansion of the wave function on a predefined basis set. However, these methods typically require an extensive basis set to accurately recover the electron correlation, leading to an explosion of computational degrees of freedom \cite{Sherrill1999, Crawford2000} with the number of electrons. 
Machine learning methods utilize deep neural networks to construct non-linear approximations and explicitly embed electron correlations into the wave function,  including DeepWF \cite{Han2020}, PauliNet \cite{Hermann2020}, FermiNet \cite{Pfau2020}, LapNet \cite{li2024computational}, and their extensions \cite{Hermann2023, pfau2024excited, li2024spin}. By integrating with the variational Monte Carlo framework \cite{ceperley1977monte, bressanini1999between}, these methods have demonstrated the superior expressivity to wave functions and yielded exceptional performance, which establishes a new approach for ab initio calculations.

While Monte Carlo based machine learning methods have achieved remarkable success, a deterministic framework remains necessary for applications requiring noise-free energy landscapes \cite{Foulkes2001}, dense excited-state spectra \cite{Chan2011}, and stable coherent dynamics \cite{Paeckel2019}.
The sum-of-Gaussians tensor neural network (SOG-TNN) has emerged as a deterministic alternative \cite{Zhou2025SOG}, which utilizes deep neural networks to dynamically construct a non-linear basis set for approximating the high-dimensional many-electron wave function.
The TNN representation has been shown to be promising in solving a diverse range of high-dimensional problems \cite{Wang2024Tensor,wang2024multi,Yu2026Tensor,Hu2024Tackling,Lin2025Tensor,Lin2026Solving,Wang2025Tensor}. 
By introducing the SOG decomposition and range splitting strategy to efficiently resolve the Coulomb interaction, the SOG-TNN successfully transforms the high-dimensional energy integrals into tractable one-dimensional evaluations.
The Slater determinant \cite{slater1929theory} can be additionally employed such that the SOG-TNN satisfies the antisymmetry constrains for fermionic systems \cite{Zhou2025SOG}. 
The attractive performance of the  approach in accuracy and efficiency has been demonstrated \cite{Zhou2025SOG}, but given its role as a non-linear neural network ansatz, a comprehensive analysis of its convergence behavior and low-rank expressivity in large-scale many-electron systems remains of significance.

In this paper, we investigate the convergence behavior and approximation capabilities of the SOG-TNN using the one-dimensional (1D) soft-Coulomb system, which is a widely recognized benchmark with strong electron correlations and complex many-body dynamics \cite{Wagner2012, Zhou2024, Coe2011, Stoudenmire2012}.
We introduce the weighted balanced truncation (WBT) method \cite{lin2025weighted} to reduce the number of Gaussians in the SOG factorization. Interestingly, for the soft-Coulomb kernel, the model reduction not only significantly reduces the number of Gaussians but also ensures that the Chebyshev expansion is efficient enough to well decompose all the Gaussians into tensor representations. 
Due to these advantages, the computational cost of the SOG-TNN is dramatically reduced under the same level of precision. This allows us to explore the convergence of the approach with a large number of electrons on a single GPU card. 
For fermionic systems ranging from hydrogen to oxygen atoms, the SOG-TNN yields ground-state energies with an accuracy several digits higher than chemical accuracy under a remarkably compressed basis size of $P \le 100$.  Furthermore, spectral convergence with respect to the basis size $P$ is observed for the broad range of elements up to oxygen. These findings highlight the promising capability of SOG-TNN to approximate the strong electron correlation while drastically mitigating basis set truncation errors, revealing that the approach is superior for high-fidelity quantum calculations of complex many-electron systems across diverse physical and chemical regimes \cite{Orus2014,Cassella2023,Haegeman2011}.

\section{Method}
\label{sec::method}

Consider the 1D soft-Coulomb system with $N$ electrons and $M$ ion under the Born-Oppenheimer approximation \cite{Born1927BO}, described by the  Schr\"{o}dinger equation, $\mathcal{H}\Psi=E\Psi$, where $E$ is the eigenvalue, and the Hamiltonian $\mathcal{H}$ is defined by,
\begin{equation}
\label{Hamilton}
\begin{gathered}
    \mathcal{H} := \sum_{i=1}^{N}\left(-\frac{1}{2}\nabla_{r_i}^2 -\sum_{J=1}^{M} z_Jv(|r_i-R_J|) + \sum_{j=i+1}^Nv(r_{ij})\right)\\
     + \sum_{I<J} z_Iz_Jv(R_{IJ}),  ~~~~~~~~~~~~~~~~~~~~~~~~~~~~~~~~~~~~~~~ 
\end{gathered}
\end{equation}
with $v(u) = 1/\sqrt{1+u^2}$ being the kernel of the soft-Coulomb interaction. In Eq.~\eqref{Hamilton}, $r_i$ and $R_J$ denote the spatial coordinates of the $i$th electron and $J$th ion, $z_J$ is the valence, and $r_{ij}$ and $R_{IJ}$ are interparticle distances. Let $V_\mathcal{A}$  in $\mathbb{R}^{N}$ denote the function space that satisfies the Pauli exclusion principle, i.e., the action of the exchange operator $T_{ij}$ that commutes $r_i$ and $r_j$ satisfies
    \begin{equation}
    \label{eq::Pauli}
    T_{ij}\Psi=-\Psi,\quad 1\le i\neq j\le N.
\end{equation}
The ground state energy $E_0$ can be solved by the minimization problem, 
\begin{equation} 
\label{SE}
E_{0}=\operatorname*{min}_{\Psi\in V_{\mathcal{A}}}{\frac{\langle\Psi|{\mathcal{H}}|\Psi\rangle}{\langle\Psi|\Psi\rangle}}.
\end{equation}
Let $\Psi_0$ be the corresponding eigenfunction of the lowest eigenvalue. The exponential decay property \cite{Agmon1982} of the wave function allows the spatial domain to be truncated to a finite domain $\Omega^N = [-r_c, r_c]^N$. 
In the following, we present the SOG-TNN framework and the model reduction technique for the optimal number of Gaussians of the SOG factorization.

\subsection{Tensor neural network representation}
\label{sec::SOGTNN_framework}
Direct minimization of Eq.~\eqref{SE} is computationally intractable due to the high-dimensional integral and the anti-symmetry restriction. 
The TNN is a recent framework for the solution of the curse of dimensionality by using tensor representation of approximate bases. 
Zhou {\it et al.} \cite{Zhou2025SOG} first proposed to approximate the many-electron wave function by the TNN with the anti-symmetric 
property preserved by employing the Slater determinant \cite{slater1929theory}, such that
\begin{equation}
\label{SlaterTNN}
\begin{gathered}
\Psi(\bm r;\mathbf{\Theta}) = \sum_{p=1}^P \alpha_p \mathcal{A}\left(\bm{\Phi}_p^\uparrow\left(\bm r^\uparrow;\mathbf{\Theta}^{\uparrow}\right)\right) \cdot \mathcal{A}\left(\bm{\Phi}_p^\downarrow\left(\bm r^\downarrow;\mathbf{\Theta}^{\downarrow}\right)\right),
\end{gathered}
\end{equation}
where $P$ is the basis size, $\{\alpha_p\}$ are linear combination coefficients, $\mathbf{\Theta}=[\mathbf{\Theta}^{\uparrow}, \mathbf{\Theta}^{\downarrow}]$ are trainable parameters of neutral networks, and $\mathcal{A}(\cdot)$ denotes the anti-symmetrization operator acting on the $p$th basis vector $\bm{\Phi}_p=(\phi_{p,1},\cdots,\phi_{p,N})^{\top}$. 
Let $N^\uparrow = \lceil N/2 \rceil$ and $N^\downarrow = N - N^\uparrow$ denote the number of electrons with spin-up and spin-down states, respectively, and $\tau^\uparrow(i)$ and $\tau^\downarrow(j)$ represent the real indices of electrons of the $i$th spin-up and $j$th spin-down electrons.
The anti-symmetrization operator is expressed as,
\begin{equation}
   \label{eq::anti-ansatz}
   \begin{aligned}   &\mathcal{A}\left(\bm{\Phi}_p^*\left(\mathbf{r}^*;\mathbf{\Theta}^{*}\right)\right)
       :={\frac{1}{N^{*}!}}\\&\cdot{\left|\begin{array}{ccc}{\phi_{p,\tau^{*}(1)}\left(r_{\tau^{*}(1)}\right)}&{\cdots}&{\phi_{p,\tau^{*}(1)}\left(r_{\tau^{*}(N^{*})}\right)}\\{\vdots}&{\ddots}&{\vdots}\\{\phi_{p,\tau^{*}(N^{*})}\left(r_{\tau^{*}(1)}\right)}&{\cdots}&{\phi_{p,\tau^{*}(N^{*})}\left(r_{\tau^{*}(N^{*})}\right)}\end{array}\right|},
   \end{aligned}
\end{equation} 
where `$*$' can be `$\uparrow$' or `$\downarrow$' to represent the spin-up or spin-down case. In the TNN framework, the basis function is parameterized using the tensor product of 1D neural network orbitals. Let $\bm{\phi}_i(x_i; \bm{\theta}_i) = \left(\phi_{1,i}, \dots, \phi _{P,i}\right)^{\top}$ be the collection of the $i$th component of the $P$ bases. It is described by
\begin{equation}\label{eq::FNN_basis}
\begin{gathered}
\bm{\phi}_{i}(x;\bm{\theta}_i)=\bigl[\bm{F}_{i}^{(l_{\text{max}})}\circ\bm{F}_{i}^{(l_{\text{max}}-1)}
\circ \cdots \circ \bm{F}_{i}^{(1)}\bigr](x), \\
\bm{F}_{i}^{(l)}(\bm{X})= \bm{\eth}^{(l)} \bigl(\bm{W}_{i}^{(l)}\bm{X} 
+ \bm{b}_{i}^{(l)}\bigr),
\end{gathered}
\end{equation}
where $\bm{\eth}$ is the element-wise activation function, and $\bm{\theta}_i$ encapsulates the trainable weights $\bm{W}_i^{(l)} \in \mathbb{R}^{N_l \times N_{l-1}}$ and biases $\bm{b}_i^{(l)} \in \mathbb{R}^{N_l}$ for a network architecture of $[N_0, N_1, \dots, N_{l_{\text{max}}}]$ neurons. Here, $N_0=1$ and $N_{l_{\text{max}}}=P$ are fixed numbers.

By Eqs.~\eqref{SlaterTNN}-\eqref{eq::FNN_basis}, the energy functional Eq.~\eqref{SE} can be deterministically evaluated in each dimension. The minimization of Eq.~\eqref{SE} is performed by the 
least-square method. Let $\left\langle\Psi|\Psi\right\rangle_{p_1,p_2}$ represent the inner product of the $p_1$th and $p_2$th bases in Eq. \eqref{SlaterTNN}, expressed by,
\begin{equation}\label{eq::inner_product} 
\left\langle\Psi|\Psi\right\rangle_{p_1,p_2}=\frac{1}{N^{\uparrow}! \cdot N^{\downarrow}!}\left|\bm M_{p_1,p_2}^{\uparrow}\right| \cdot \left|\bm M_{p_1,p_2}^{\downarrow}\right|,  
\end{equation}
where $\bm M_{p_1,p_2}^*$ can be calculated efficiently by Löwdin's rule \cite{Lowdin1955}
\begin{equation}\label{eq::lowdin}
\begin{gathered}
\bm M^*(\bm{\Phi},\bm{\widetilde{\Phi}})=\begin{bmatrix}
\langle\phi_{1} | \widetilde{\phi}_{1} \rangle  & \dots &\langle\phi_{1} | \widetilde{\phi}_{n} \rangle \\
 \vdots & \ddots  & \vdots\\
\langle\phi_{n} | \widetilde{\phi}_{1} \rangle  & \dots &\langle\phi_{n} | \widetilde{\phi}_{n} \rangle
\end{bmatrix}.
\end{gathered}
\end{equation}

Combing with the technique of the low-rank update, one can define the biorthogonalization $\bm{\Xi}_{p_1,p_2}^{*} \in \mathbb{R}^{N^{*}}$ \cite{beylkin2008approximating} by
\begin{equation}\label{eq::biorthogonalization}
\begin{gathered}
\bm{\Xi}_{p_1,p_2}^{*} = (\bm{M}_{p_1,p_2}^{*})^{-1}\bm{\Phi}_{p_1}^{*}.
\end{gathered}
\end{equation}
Let $\xi_{p_1,p_2,\tau^{*}(i)}$ represent the $i$th element of the biorthogonalization. The stiffness matrix and ion-electron potential matrix in Eq.~\eqref{SE} can be fast evaluated in an element-wise manner by
\begin{equation}\label{eq::Slater_all}
\begin{gathered} 
\left\langle\nabla_i\Psi|\nabla_i\Psi\right\rangle_{p_1,p_2}=\left\langle\Psi|\Psi\right\rangle_{p_1,p_2}\left\langle\nabla_i\xi_{p_1,p_2,i}|\nabla_i\phi_{p_2,i}\right\rangle,\\
\left\langle\Psi\left|v_{\text{ie}}\right|\Psi\right\rangle_{p_1,p_2}=\left\langle\Psi|\Psi\right\rangle_{p_1,p_2} \left\langle\xi_{p_1,p_2,i}\left|v_{\text{ie}}\right|\phi_{p_2,i}\right\rangle, 
\end{gathered}
\end{equation}
where $v_{\text{ie}}=-\sum_{J=1}^{M} z_Jv(|r_i-R_J|)$ denotes the ion-electron potential.
For electron-electron interactions, the normalized matrix element of the $i$th and $j$th electrons with parallel spins is given by
\begin{equation}\label{eq::Slater_ee_parallel}
\begin{aligned}
&\frac{\left\langle\Psi\left|v\left(r_{ij}\right)\right|\Psi\right\rangle_{p_1,p_2}}{\left\langle\Psi|\Psi\right\rangle_{p_1,p_2}} \\ 
=&\left\langle 
\begin{vmatrix}
\xi_{p_1,p_2,i}(r_i)  & \xi_{p_1,p_2,i}(r_j)\\
\xi_{p_1,p_2,j}(r_i)  & \xi_{p_1,p_2,j}(r_j)
\end{vmatrix}\left|v\left(r_{ij}\right)\right|\phi_{p_2,i}\phi_{p_2,j}\right\rangle.
\end{aligned}
\end{equation}
Correspondingly, the case of anti-parallel spins reads
\begin{equation}\label{eq::Slater_ee_anti-parallel}
\begin{aligned}
\frac{\left\langle\Psi\left|v\left(r_{ij}\right)\right|\Psi\right\rangle_{p_1,p_2}}{\left\langle\Psi|\Psi\right\rangle_{p_1,p_2}}
=\left\langle\xi_{p_1,p_2,i}\xi_{p_1,p_2,j}\left|v\left(r_{ij}\right)\right|\phi_{p_2,i}\phi_{p_2,j}\right\rangle.
\end{aligned}
\end{equation}

By Eqs.~\eqref{eq::inner_product}-\eqref{eq::Slater_ee_anti-parallel}, the TNN assembles the many-electron Hamiltonian via low-rank updates while preserving powerful non-linear approximations for high-dimensional wave functions. Nevertheless, the electron-electron kernel inherently lacks a separable tensor-product structure. This kernel convolution is coupled with neural network orbitals in Eq.~\eqref{eq::FNN_basis}, and its calculation is computationally expensive in both numerical quadrature and gradient backpropagation \cite{baydin2018automatic}, and thus an efficient tensor factorization of the soft-Coulomb kernel is required for evaluating the convolutions.

\subsection{Kernel tensor factorization}
\label{sec::decoup}
To efficiently evaluate the electron-electron kernel convolutions in Eqs.~\eqref{eq::Slater_ee_parallel}-\eqref{eq::Slater_ee_anti-parallel}, one essential idea is the tensor factorization of the interacting kernel.
In \cite{Zhou2025SOG}, the truncated bilinear series approximation (BSA) \cite{beylkin2005, beylkin2010} is used for a SOG approximation of the Coulomb kernel and thus each Gaussian can be factorized into a tensor product. The SOG is a technique of kernel approximation to design fast algorithms for kernel convolution \cite{gan2025fast,greengard2018anisotropic} and particle interactions \cite{predescu2020,RBSOG,Gao2026Fast,ji2025machine,Ji:2026JCP}. The BSA for the soft-Coulomb kernel is,
\begin{equation}\label{eq::SOG}
    v(u) \approx \sum_{\ell=-\infty}^{\infty} \omega_\ell \, e^{-s_\ell u^2},
\end{equation}
where $s_\ell = 1/(2b^{2\ell}\sigma^2)$ and $\omega_\ell = (2\ln b \cdot e^{-s_\ell})/(\sqrt{2\pi}b^\ell\sigma)$ with $b>1$ and $\sigma>0$ being parameters of tuning the error:
\begin{align}
\left|1 - \frac{1}{v(u)}\sum_{\ell=-\infty}^{\infty} \omega_\ell \, e^{-s_\ell u^2}\right|
\lesssim 2\sqrt{2}\exp\!\left(-\frac{\pi^2}{2\ln b}\right).\
\label{eq::sog_error}
\end{align}
The BSA expansion is also known as the u-series \cite{predescu2020} in molecular dynamics due to the asymptotic convergence as parameter $b\rightarrow 1$.

The computational complexity of electron-electron kernel convolution scales linearly with the number of Gaussians. To improve the efficiency of the SOG-TNN, we introduce two 
model reduction steps to reduce both the number of Gaussians and the order of Chebyshev expansion. 

In the first step, we introduce the weighted balanced truncation method \cite{lin2025weighted} to effectively compress the required number of Gaussians, such that it holds
\begin{equation}\label{eq::WBT}
\left| \sum_{\ell=-L_1}^{L_2} \omega_\ell e^{-s_\ell u^2} - \sum_{\ell=1}^{L} \widetilde{\omega}_\ell e^{-\widetilde{s}_\ell u^2} \right| < \epsilon
\end{equation}
for $L$ being much smaller than  $L_1+L_2$ under a given error tolerance $\epsilon$. Note that the error estimate of the truncated BSA can be found in \cite{LIANG2025101759}.
The balance truncation \cite{moore1981principal,antoulas2001approximation,gao2022kernel} is a classical technique for the model reduction, which transforms the SOG into a sum of poles, writes the resultant sum into a matrix form,
and then introduces the Hankel singular decomposition to reduce the number of poles.
The WBT introduces a customized weight function into the balanced truncation method such that number $L$ can be optimized for approximation within a given domain.  

The second step lies in the tensor factorization of Gaussians. When the Gaussian bandwidth is large,  the Chebyshev polynomial expansion is employed to approximate the $\ell$th Gaussian into a separable tensor-product structure,
\begin{equation}
\label{eq:chebyshev_expansion}
  e^{-\widetilde{s}_\ell (r-r')^2}\approx \sum_{m+n\le S} C_{m,n}^{\ell} T_m\left(\frac{r}{r_c}\right)T_n\left(\frac{r'}{r_c}\right),
\end{equation}
where $T_n(\cdot)$ denotes the $n$th Chebyshev polynomial, and coefficients ${C}^{\ell}_{m,n}$ can be precomputed by the orthogonal projection. 
The truncation order $S$ can be explicitly determined by the factorially decaying error bound \cite{chebyshev}.
Let us define the Chebyshev coefficient matrix, $\bm C^{\ell}$, which is $(S+1)$-by-$(S+1)$ with the $(m,n)$-element being $C_{m,n}^{\ell}$ for $m+n\leq S$ and zero otherwise.
We apply singular value decomposition (SVD) to the matrix $\bm{C}^{\ell}$ for further model reduction. 
By discarding singular components below the target tolerance $\epsilon$, the matrices are compressed such that their matrix ranks $K^{\ell} \ll S$. 

By the two procedures, the evaluation of the electron-electron kernel convolutions in Eqs.~\eqref{eq::Slater_ee_parallel} and \eqref{eq::Slater_ee_anti-parallel} are effectively reduced to $L$ terms, where each one is a $K^{\ell}$-term summation of 1D integrals. Let $K_{\text{max}}$ be the maximum value of $K^{\ell}$ for $\ell=1, \cdots, L$.
Table~\ref{tab::params_SOG} illustrates the results of the model reductions for the soft-Coulomb kernel under the error tolerance of $\epsilon=10^{-12}$ with corresponding SOG parameters $\sigma=1$ and $b=1.187800189600461$.
One can clearly observe the significant effect of the model reductions. The BSA requires $\sim220$ Gaussians for the desired accuracy, and the WBT successfully compresses the number into $26\sim28$ terms 
for different $r_c$. This is nearly an order of magnitude in reduction. Moreover, the subsequent SVD further reduces the Chebyshev expansion rank by approximately two-thirds.

\begin{table}[htbp]
\caption{ Results for kernel tensor factorization at different $r_c$ under error tolerance $\epsilon=10^{-12}$}
\label{tab::params_SOG}
\renewcommand{\arraystretch}{1.4} 
\begin{ruledtabular} 
\begin{tabular*}{\columnwidth}{@{\extracolsep{\fill}}cccccc@{}}
$r_c$ (a.u.) & $L_1$ & $L_2$ & $L$ & $S$ & $ K_{\text{max}}$ \\ \hline
13    & 12 & 210 & 26 & 656 & 229 \\
15    & 12 & 212 & 27 & 784 & 270 \\
20    & 12 & 216 & 28 & 1124 & 361 \\
\end{tabular*}
\end{ruledtabular}
\end{table}

The tensor factorization and model reduction procedure results in that the electron-electron kernel convolution can be evaluated by 1D integrals, substantially maintaining the separable structure of
all variables in the network architecture, a feature specifically tailored for the TNN framework to enhance its training efficiency. By leveraging the Gauss-Legendre quadrature, the SOG-TNN achieves a high-precision assembly of the global Hamiltonian in Eq.~\eqref{Hamilton} with a complexity of $\mathcal{O}(P^2N^3LK_{\text{max}}T_{\text{1D}})$ operations, where the $T_{\text{1D}}$ denotes the computational complexity of 1D numerical integration.

\section{Results}
\label{sec::num_result}

In this section, we demonstrate the expressivity of the SOG-TNN method across a variety of atomic systems. Numerical solutions of the Schr\"{o}{dinger equation for 1D soft-Coulomb systems ranging from H atom to O atom are performed. In the results, we will evaluate the learning stability of the SOG-TNN, the rapid decay characteristics of basis functions, and the enhanced computational efficiency in comparison with methods of conventional basis sets. The accuracy of the ground state energy reported in these results is measured by the relative error $\mathcal{E}_r$ defined by
$\mathcal{E}_r=(E-E_0)/|E_0|$, where $E_0$ is the reference result calculated by specific methods with sufficient precision. These calculations are performed on a single NVIDIA RTX 4090D GPU (24 GB memory, 384-bit interface, 1040 GB/s bandwidth).
We employ a configuration interaction (CI) method based on Legendre-Shen orthogonal polynomials \cite{Shen2016Efficient} for comparative baseline calculations. The CI method incorporates a sparse grid (SG) truncation for acceleration. The SG exploits sparsity patterns derived from the mixed-derivative regularity analysis of the wave function~\cite{yserentant2004regularity,Yserentant2005Sparse,Yserentant2007Hyperbolic,Yserentant2012Mixed}, thereby drastically reducing the required degrees of freedom while preserving high convergence rates \cite{griebel2007sparse,griebel2009tensor}. 

\subsection{Stable learning of SOG-TNN}
\label{sec::stable_learning}

\begin{figure*}[t] 
\centering   \includegraphics[width=0.9\textwidth]{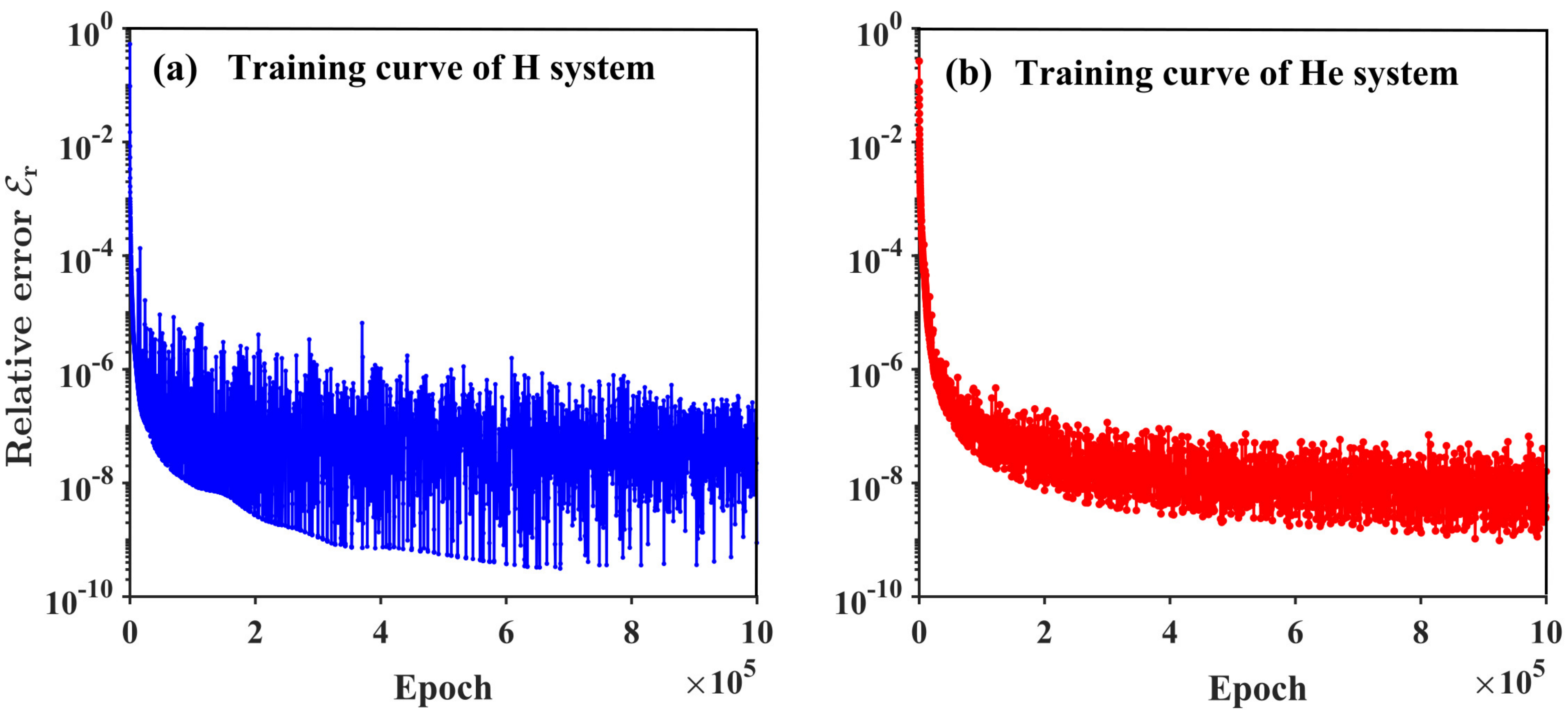}
\caption{Training curves of error $\mathcal{E}_r$ for (a) H and (b) He with basis size $P=12$ and $P=20$, respectively.}
\label{fig::Stability}
\end{figure*}

The ground-state energy represents the minimum energy attainable as the wave function traverses the subspace $V_{\mathcal{A}}$. Consequently, if $\mathcal{E}_r$ remains consistently positive throughout the learning process, it indicates that the method is numerically stable; otherwise, the method converges toward non-physical results. We apply the SOG-TNN method to H and He systems, for which the SG-CI method exclusively provides high-precision reference ground-state energies: $E^{\text{H}}_0 = -0.6697771382138$ with $r_c = 15$ a.u., and $E^{\text{He}}_0 = -2.23825782410$ with $r_c = 13$ a.u.. The standard Adam optimizer \cite{kingma2014adam} is employed with fixed learning rate $\eta=10^{-3}$. 
Fig.~\ref{fig::Stability} illustrates the evolution of the solution error with epochs. Both H and He systems exhibit a rapid descent of energy error to $10^{-8}$ during the initial phase, while maintaining a consistently positive $\mathcal{E}_r$ throughout the process. These results demonstrate that the training process of the SOG-TNN method is remarkably stable, consistently yielding high-precision solutions that possess clear physical significance.

While the constant learning rate Adam optimizer achieves the high precision, stochastic oscillations does lead to slow convergence. In the following calculations, we design a three-phase hybrid optimization for the learning strategy. Specifically, the optimization starts with the RAdam optimizer \cite{liu2019variance} for $6.35\times 10^{5}$ epochs, employing a cosine annealing scheduler with warm restarts ($T_0=5000$, $T_{\text{mult}}=2$) \cite{loshchilov2016sgdr} to periodically anneal the learning rate from $\eta= 10^{-3}$ to $10^{-5}$. Subsequently, resuming from the best parameter $\bm{\Theta}$ in Phase 1, a non-restarting cosine annealing phase is executed for $1.6\times 10^{5}$ epochs, smoothly decaying the learning rate from $\eta=10^{-4}$ to $10^{-7}$ to ensure stable convergence. Ultimately, we turn to the pseudo-second-order L-BFGS optimizer \cite{liu1989limited} for $5\times 10^{3}$ epochs with a learning rate of $\eta=0.01$ for fine-tuning convergence. In the subsequent convergence tests, we adopt the hybrid optimization strategy to achieve numerical solutions of high accuracy.

\subsection{Convergence rate with basis size}
\label{sec::conv_analysis}

\begin{table*}[htbp]
\caption{Computational parameters, the reference energy $E_0$ and the SOG-TNN results}
\label{tab:params_energies_jcp}
\renewcommand{\arraystretch}{1.4} 
\begin{ruledtabular} 
\begin{tabular*}{\textwidth}{@{\extracolsep{\fill}}ccccccc@{}}
System & $r_c$ (a.u.) & TNN sizes & Basis size $P$ & $E$  & $E_{0}$ & $\Delta E$  \\ \hline
H  & 15 & $[1, 64, 128, p]$      & 12 & -0.6697771382065866   & -0.6697771382138     & $\bf{1.08E-11}$\\
He & 13 & $[1, 64, 128, p]$      & 20 & -2.2382578223930367   & -2.2382578241080     & $\bf{7.66E-10}$\\
Li & 20 & $[1, 64, 128, p]$      & 40 & -4.210531645922184    & -4.210531647613249   & $\bf{4.02E-10}$\\
Be & 20 & $[1, 64, 128, p]$      & 60 & -6.785077942563639    & -6.78507795170852    & $\bf{1.35E-9}$\\
B  & 20 & $[1, 64, 128, p]$      & 64 & -9.819712286746151    & -9.81971246294425    & $\bf{1.79E-8}$\\
C  & 20 & $[1, 64, 128, 256, p]$ & 80 & -13.331542394279513   & -13.33154314459545   & $\bf{5.63E-8}$\\
N  & 20 & $[1, 64, 128, 256, p]$ & 88 & -17.29164413981273    & -17.29164709368697   & $\bf{1.71E-7}$\\
O  & 20 & $[1, 64, 128, 256, p]$ & 84 & -21.699553846182102   & -21.69957056180725  & $\bf{7.70E-7}$\\
\end{tabular*}
\end{ruledtabular}
\end{table*}

\begin{figure*}[htbp] 
\centering   
\includegraphics[width=0.9\textwidth]{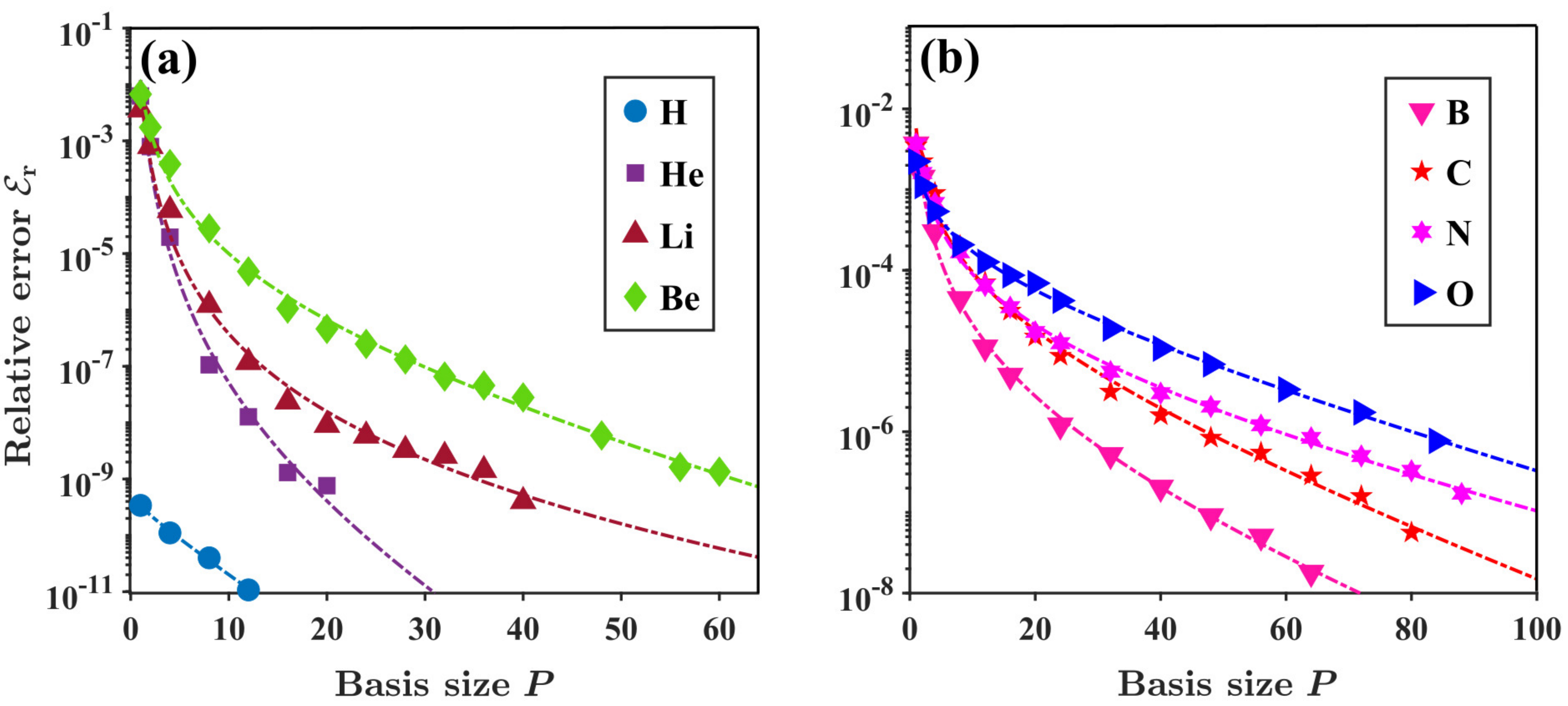}
\caption{Convergence of relative error $\mathcal{E}_r$ as function of the basis size $P$ for 1D soft-Coulomb systems from H to O.  
The dashed lines are fitting lines by ansatz $\mathcal{E}_r=C\cdot P^{-\beta}\cdot e^{-\gamma P}$.} 
\label{fig::Convergence} 
\end{figure*}

\begin{table}[htbp]
\caption{Fitting parameters ($R^2$, $C$, $\beta$, and $\gamma$) of the curves in Fig.~\ref{fig::Convergence}}
\label{tab::fitting_params}
\renewcommand{\arraystretch}{1.4} 
\begin{ruledtabular} 
\begin{tabular*}{\columnwidth}{@{\extracolsep{\fill}}ccccc@{}}
System & $R^2$ & $C$ & $\beta$ & $\gamma$ \\ \hline
H  & 1.00  & 2.95E-10 & 0.154 & 0.273 \\
He & 0.996 & 0.0314   & 4.78  & 0.152 \\
Li & 0.989 & 0.0365   & 4.26  & 0.0226 \\
Be & 0.996 & 0.0781   & 2.72  & 0.0812 \\
B  & 0.998 & 0.0535   & 2.17  & 0.0551 \\
C  & 0.994 & 0.0811   & 1.56  & 0.0573 \\
N  & 0.997 & 0.0860   & 1.60  & 0.0346 \\
O  & 0.999 & 0.0501   & 0.933 & 0.0457 \\
\end{tabular*}
\end{ruledtabular}
\end{table}

The SOG-TNN is a high-precision solver and allows for the attainment of pre-specified level of accuracy through systematic parameter tuning. 
This characteristic facilitates rigorous numerical experiments to evaluate the convergence rate with respect to the parameters of interest. 
We now assess the approximation capacity of the SOG-TNN by evaluating the accuracy $\mathcal{E}_r$ against the basis size $P$. 
Table~\ref{tab:params_energies_jcp} gives the network parameters, the reference solution, and the SOG-TNN results. 
Fig.~\ref{fig::Convergence} presents the results for systems ranging from H to O with the increase of $P$. 
From these results, one can clearly observe that the SOG-TNN can achieve high-precision approximations using only a small basis number $P$.
Specifically, the chemical accuracy, corresponding to an absolute error within $1.6 \times 10^{-3}$ a.u., is successfully attained across all tested systems with a highly compact basis size of $P \le 20$.
As the basis size is expanded to $P=60$, the errors are smaller than $10^{-7}$ for atoms H, He, Li, Be and B, and smaller than $10^{-5}$ for C, N and O.
Furthermore, the error curves exhibit a spectral-like decay rate with respect to $P$. 
To quantitatively analyze this convergence behavior, we fit the following error ansatz:
\begin{equation}\label{eq::convergence_model}
   \mathcal{E}_r=C\cdot P^{-\beta}\cdot e^{-\gamma P},
\end{equation}
where $C$, $\beta$ and $\gamma$ are constants independent of $P$. The fitting parameters for the curves in Fig.~\ref{fig::Convergence} 
are displayed in Table~\ref{tab::fitting_params}.  
These numerical results clearly validate the ansatz, with coefficients of determination $R^2\ge 0.989$ for all systems from H to O atom. 
It should be noted that, for systems lacking explicit reference energies (from Li to O), the exact asymptotic limits $E_{0}$ are simultaneously extrapolated by admitting the ansatz Eq.~\eqref{eq::convergence_model}. The non-monotonic downward trend exhibited by the exponential decay rate ensures that 
the SOG-TNN remains capable of achieving efficient solutions using a small basis size under the situations of
multi-GPU parallel computing for larger systems.

\subsection{Comparison with the SG-CI results}
\label{sec::comp_SG}

The SOG-TNN fully exploits the powerful approximation capabilities of neural networks, enabling the high-precision solution of the Schr\"{o}dinger equation using an exceptionally small basis size $P$. 
We demonstrate this significant advantage through a comparative analysis with the results of the SG-CI method \cite{Shen2016Efficient}. 
The SG-CI successfully utilizes wave function regularity to truncate the basis size of the configuration interaction method, 
and the SOG-TNN takes advantage of neural networks to dynamically train a set if low-rank bases. Due to this similar property, 
the SG-CI is a suitable baseline for the efficiency comparison. Fig.~\ref{fig::Comparison} illustrates the error $\mathcal{E}_r$ 
as function of the basis size $P$ for both methods across the systems from H to O. 
One observes that the SOG-TNN achieves much higher accuracy for fixed basis size.  
For example, the SOG-TNN  for the Li system has $10^{-8}$ accuracy with a mere basis size $P=20$, whereas the SG-CI method demands about $5\times 10^{4}$ bases to achieve the accuracy.
For multi-electron systems, the SG-CI shows minimal error improvement with basis expansion, but the SOG-TNN captures correlations effectively, 
achieving spectral-like convergence.
For example, for the C system, the SOG-TNN attains  $10^{-7}$ accuracy with a highly compact basis of $P=80$, 
whereas the SG-CI approach struggles to go beyond $10^{-2}$ accuracy in spite that the basis size has expanded to tens of thousands.
These results show that, by employing nonlinear neural network bases, the SOG-TNN is capable of dynamically identifying efficient representations 
for complex many-electron wave functions. This demonstrates the robustness of the neural network framework to 
overcome the bottleneck of deterministic methods in accurate solutions for the high-dimensional Schr\"{o}dinger equation.

\begin{figure*}[htbp] 
\centering   
\includegraphics[width=0.9\textwidth]{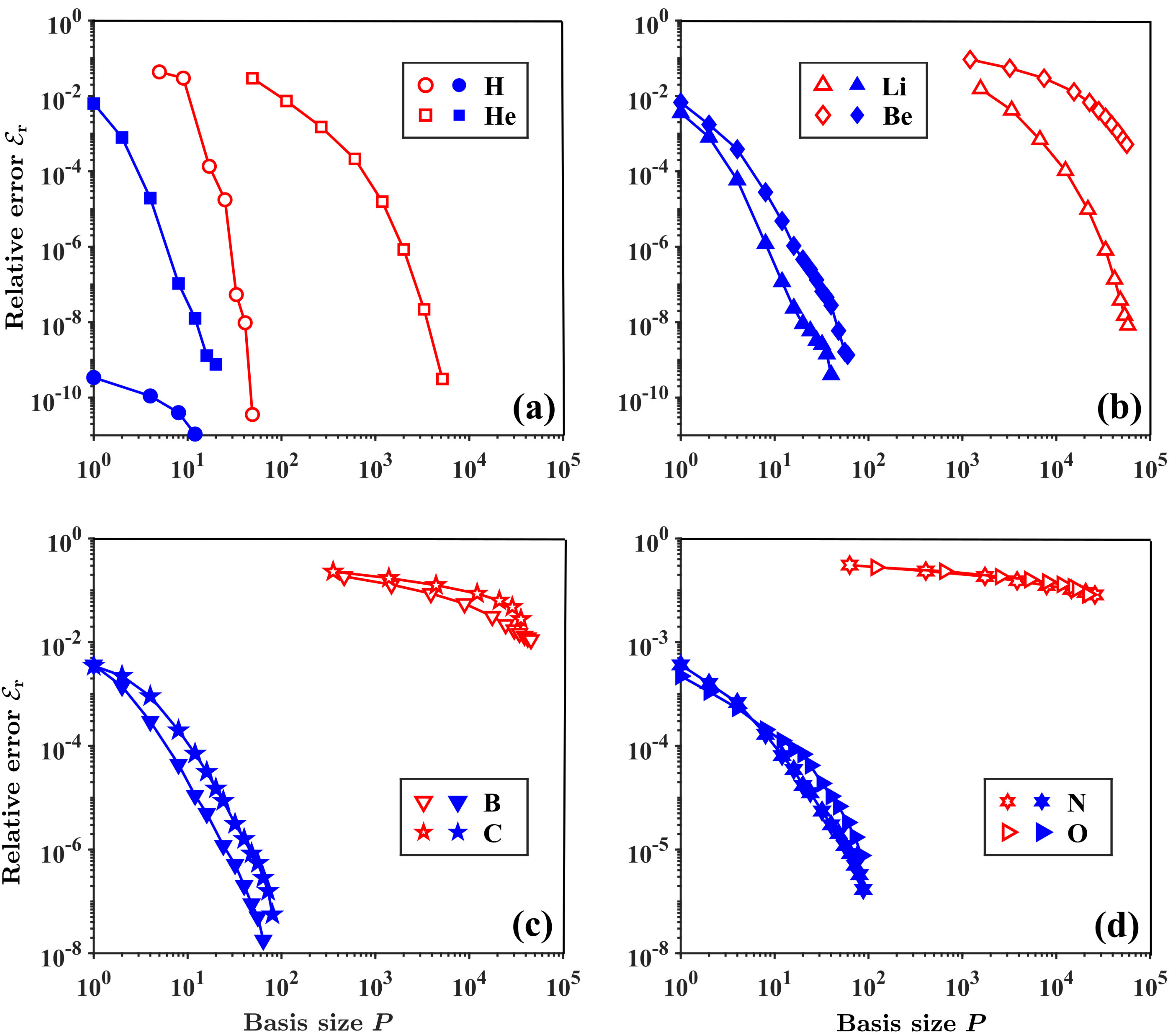}
\caption{Log-log convergence of $\mathcal{E}_r$ against the basis size $P$ for the SOG-TNN (blue color) and SG-CI methods (red color). (a) H and He atoms; (b) Li and Be atoms; (c)  B and C atoms;  
and (d) N and O atoms.}
\label{fig::Comparison} 
\end{figure*}

\section{Conclusion}
\label{sec::conclusion}

In this work, we proposed the SOG-TNN as an ultra-efficient deterministic solver to the many-electron Schr\"{o}dinger equation for 1D soft-Coulomb systems.
By using Slater determinant to preserve the anti-symmetry of wave function, we combine model reduction techniques with the SOG decomposition for 
a low-rank tensor factorization of the soft-Coulomb kernel, resulting in highly efficient SOG-TNN for approximating the ground-state solution. 
Numerical results demonstrate robust spectral convergence of the approach with respect to the basis number $P$,
and high accuracy of the solution with a small $P$. The SOG-TNN is shown promising for the high-dimensional problem. 
Our future work will focus on extending the SOG-TNN architecture to more complex many-electron systems, particularly, in applications of excited-state calculations and time-dependent quantum dynamics.

\section*{Acknowledgement}

The work of Q. Zhou and Z. Xu is supported by National Natural Science Foundation of China (grants No. 12426304, 12325113 and 125B2023). The work of H. Xie is supported by the Strategic Priority Research Program of the Chinese Academy of Sciences (XDB0620203, XDB0640000, XDB0640300),  National Key Laboratory of Computational Physics (grant No. 6142A05230501), 
 National Key Research and Development Program of China (2023YFB3309104), 
National Natural Science Foundation of China (grant No. 12331015), 
and Science Challenge Project of National Center for Mathematics and Interdisciplinary Science, CAS (grant No. TZ2024009). The work is partially supported by the SJTU Kunpeng \& Ascend Center of Excellence.

\section*{Conflict of interest}

The authors declare that they have no conflict of interest.

\section*{Data Availability Statement}

The data that support the findings of this study are available from the corresponding author upon reasonable request.

\end{document}